\documentclass[12pt]{article}
\RequirePackage[l2tabu, orthodox]{nag}
\usepackage[dvipdfmx]{graphicx}

\usepackage{color,amsmath,slashed,booktabs,braket,hyperref,bm,authblk,graphicx,amssymb}
\usepackage[a4paper, total={16.5cm,22cm}]{geometry}

\usepackage{newtxtext}
\usepackage{newtxmath} 
\usepackage[utf8x]{inputenc} 

\usepackage[sort&compress,numbers, merge]{natbib}

\hypersetup{
  colorlinks = true,
  linkcolor = blue,
  citecolor = blue,
}

\title{{\bfseries 
Constraining FIMP from the structure formation \\ 
of the Universe: analytic mapping from $m_{\mathrm{WDM}}$
}}
\author[1]{Ayuki Kamada}
\affil[1]{{\small \textit{Center for Theoretical Physics of the Universe, Institute for Basic Science (IBS),  Daejeon 34126, Korea}}}
\author[2]{Keisuke Yanagi}
\affil[2]{{\small \textit{Department of Physics, University of Tokyo, Tokyo 113--0033,Japan}}}
\date{}

\newcommand{\vev}[1]{\left\langle #1\right\rangle}

\newcommand\Order{\mathop{\mathcal{O}}}

\newcommand\unit[1]{\,\mathrm{#1}}

\newcommand\keV{\unit{keV}}

\newcommand{\hc}{\mathrm{h.c.}}

\begin{document}
\hspace*{\fill} CTPU-PTC-19-21 \\
\hspace*{\fill} UT-19-16

{\let\newpage\relax\maketitle}

\begin{abstract}
A feebly interacting massive particle (FIMP), contrasting with a weakly interacting massive particle (WIMP), is an intriguing dark matter (DM) candidate.
Light (keV-scale) FIMP DM is of particular interest: its radiative decay leaves a line signal in x-ray spectra; and it is warm dark matter (WDM) and alters the galactic-scale structure formation of the Universe from that with WIMP DM.
Once a possible x-ray line is reported (e.g., $3.5 \keV$ line and $7 \keV$ FIMP DM is inferred), one has to check whether or not this FIMP DM is compatible with the structure formation.
Here is an issue: the structure formation constraint on WDM is often reported in terms of the so-called thermal WDM mass $m_{\mathrm{WDM}}$, which cannot be directly applied to FIMP parameters.
In this paper, we introduce a benchmark FIMP model that represents well a broad class of FIMP models.
A big advantage of this benchmark is that we can derive the analytic formula of the non-thermal phase space distribution of FIMPs produced from freeze-in processes.
By further deriving a certain ``warmness'' quantity, we can analytically map $m_{\mathrm{WDM}}$ to FIMP parameters.
Our analytic map indicates that $7 \keV$ FIMP DM, without entropy production or a degenerate spectrum, is in tension with the latest Lyman-$\alpha$ forest data. 
Our analytic map will be very useful for future updates of observational constraints and reports of x-ray lines.
It is also very easy to incorporate our analytic formula into a Boltzmann solver so that a linear matter power spectrum is readily accessible.
Our benchmark model will facilitate FIMP searches and particle physics model-building.
\end{abstract}

\thispagestyle{empty}
\clearpage



\section{Introduction} 
\label{sec:introduction}
Dark matter (DM) will provide important clues of physics beyond the standard model (BSM) once its particle nature becomes revealed (see Ref.~\cite{Feng:2010gw} for a review).
For past decades, a weekly interacting massive particle (WIMP) has been the most attractive candidate of DM. 
Its origin is naturally identified in electroweak-scale BSM scenarios that address the hierarchy problem, such as supersymmetric extension of the SM.
Extensive experimental and observational searches of WIMPs have been conducted (see, e.g., Refs.~\cite{Roszkowski:2017nbc, Arcadi:2017kky} for recent reviews).
Despite such efforts, however, the WIMP signal has not been reported yet.
Moreover, the Large Hadron Collider (LHC) experiment has pushed up a new-physics scale to TeV, degrading naturalness-oriented BSM scenarios and thus also WIMPs.
This situation prompts us to consider an alternative to WIMP DM more seriously. 

Feebly interacting massive particle (FIMP), which only feebly interacts with the visible sector, DM is an intriguing alternative to WIMP DM.
FIMPs are produced by the decay and/or scattering of the particles in the thermal plasma, but unlike WIMPs, they never attain the equilibrium with the plasma particles due to the feeble interaction.
This production mechanism is called \textit{freeze-in}, where FIMP production is efficient at low energy and 
thus the resultant yield is insensitive to unknown high-energy physics~\cite{Hall:2009bx} (see Ref.~\cite{Bernal:2017kxu} for a recent review).%
\footnote{We assume that the FIMP yield from decay of inflaton is negligible.
}

Light (keV-scale) FIMP DM has distinctive features.
In contrast to a WIMP, a light FIMP can be DM without any new symmetry for its stability since a small mass in combination with feeble interaction makes the FIMP sufficiently long-lived.
Moreover, rare decay of FIMP DM may be detected by x-ray searches.
Indeed, there was an report of an unidentified $3.5\keV$ line in x-ray spectra of \texttt{Chandra} and \texttt{XMM-Newton}.%
\footnote{
We refer readers to Refs.~\cite{Bulbul:2014sua, Boyarsky:2014jta} for the first two reports and Refs.~\cite{Boyarsky:2014ska, Anderson:2014tza, Iakubovskyi:2015dna, Neronov:2016wdd, Cappelluti:2017ywp} for the subsequent reports.
There have also been null-detection reports~\cite{Horiuchi:2013noa, Jeltema:2014qfa, Malyshev:2014xqa, Tamura:2014mta, Sekiya:2015jsa, Jeltema:2015mee, Aharonian:2016gzq, Tamura:2018scp, Dessert:2018qih} and claims that the DM interpretation fails to pass consistency checks from line profiles~\cite{Urban:2014yda, Carlson:2014lla}.
We note that some of the null-detection reports are just not sensitive enough to exclude the DM interpretation (see Refs.~\cite{Iakubovskyi:2015wma, Abazajian:2017tcc} for summaries) and the failure claims are refuted~\cite{Boyarsky:2014paa, Bulbul:2014ala, Ruchayskiy:2015onc, Franse:2016dln, Boyarsky:2018ktr}.
It seems not conclusive so far, while future experiments such as XQC~\cite{Figueroa-Feliciano:2015gwa} and XRISM~\cite{2018SPIE10699E..22T} will help us examine it more closely.
}
Decay of $7\keV$ FIMP DM may explain the unidentified x-ray line.
Meanwhile, such light FIMP DM is constrained by the galactic-scale structure formation of the Universe since it behaves as warm dark matter (WDM).
In particular, Lyman-$\alpha$ forest observations provide stringent constraints on WDM~\cite{Viel:2005qj, Seljak:2006qw, Viel:2006kd, Viel:2007mv, Viel:2013apy, Baur:2015jsy, Yeche:2017upn, Irsic:2017ixq}.
Observations of redshifted 21\,cm signals are also promising probes of WDM~\cite{Sitwell:2013fpa, Sekiguchi:2014wfa, Safarzadeh:2018hhg, Schneider:2018xba, Lopez-Honorez:2018ipk, Chatterjee:2019jts, Boyarsky:2019fgp}.%
\footnote{
Other used probes include the number of satellite galaxies in the Milky Way~\cite{Maccio:2009isa, Polisensky:2010rw, Lovell:2013ola, Kennedy:2013uta, Horiuchi:2013noa, Schneider:2014rda},
the delay of the reionization~\cite{Barkana:2001gr, Yoshida:2003rm, Schultz:2014eia, Lapi:2015zea, Tan:2016xvl, Lopez-Honorez:2017csg}, the counts of high-$z$ gamma-ray bursts~\cite{Mesinger:2005ah, deSouza:2013hsj}, the faint end of luminosity function of high-$z$ galaxies~\cite{Pacucci:2013jfa, Schultz:2014eia, Lapi:2015zea, Menci:2016eww, Menci:2016eui}, the flux anomaly of quadrupole lens systems~\cite{Miranda:2007rb, Inoue:2014jka, Kamada:2016vsc, Birrer:2017rpp, Gilman:2017voy, Vegetti:2018dly}.
The counts of lensed distant supernovae~\cite{Pandolfi:2014rea} and direct collapse black holes~\cite{Dayal:2017yhb} are suggested for a future use.
We also refer readers to Ref.~\cite{Governato:2002cv, Gao:2007yk, Herpich:2013yga, Governato:2014gja, Maio:2014qwa, Colin:2014sga, Gonzalez-Samaniego:2015sfp, Lovell:2016fec, Wang:2016rio, Villanueva-Domingo:2017lae, Bozek:2018ekc, Bremer:2018nuf, Fitts:2018ycl, Maccio:2019hby} for hydrodynamical simulation results differentiating WDM and CDM in galaxy formation.
}
It is very natural to ask \textit{whether or not (for instance $7\keV$) FIMP DM is compatible with these constraints}.

On the other hand, it is not easy to answer this question.
Observational constraints are often reported in terms of the thermal WDM mass $m_{\mathrm{WDM}}$.
The current strongest limits from Lyman-$\alpha$ forest data and redshifted 21\,cm signals in EDGES observations~\cite{Bowman:2018yin} are, respectively, $m_{\mathrm{WDM}} > 5.3\keV$~\cite{Irsic:2017ixq} and $m_{\mathrm{WDM}} > 6.1\keV$~\cite{Schneider:2018xba, Lopez-Honorez:2018ipk} (up to a galaxy formation model~\cite{Chatterjee:2019jts, Boyarsky:2019fgp}).
In the conventional thermal (or early decoupled) WDM model, the WDM temperature is determined such that the thermal relic abundance coincides with the observed DM mass density.
For example, for $m_{\mathrm{WDM}} = {\mathcal O}(1) \keV$, the WDM temperature is set to be $\sim 10$ times lower than the neutrino temperature.
Thus, for the equal mass, the thermal WDM is colder than FIMP DM (see below and also Ref.~\cite{Bae:2017dpt}).

In order to constrain FIMP DM, we need to repeat the following procedure~\cite{Bae:2019sby} in a model-by-model manner:
\begin{itemize}
\item[]
\begin{center}
Model $\to$ DM phase space distribution $\to$ Linear matter power spectrum $\to$ Observables.
\end{center}
\end{itemize}
It requires multidisciplinary expertise from particle physics to (computational) astrophysics. 
The last step is also time and computational resource consuming.
Therefore, it is natural to develop some map from $m_{\mathrm{WDM}}$ onto FIMP parameters.
As an example, one somehow compares the linear matter power spectra in the both models (see below and also ~\cite{Konig:2016dzg}), while omitting the last step in the above procedure.
As another example, one compares some ``warmness'' quantity that is calculable from the DM phase space distribution~\cite{Bond:1982uy, Bond:1983hb, Bond:1980ha, Davis:1981yf, Peebles:1982ib} (see below and also \cite{Kamada:2013sh}).
Then, one omits the last two steps, but still has to compute the phase space distribution of FIMPs by integrating the Collision term of the Boltzmann equation~\cite{Shaposhnikov:2006xi, Gorbunov:2008ui, Boyanovsky:2008nc, Adulpravitchai:2015mna, McDonald:2015ljz, Roland:2016gli, Heeck:2017xbu, Boulebnane:2017fxw} .

In this paper, we introduce a benchmark model of a FIMP, where the phase space distribution is analytically expressed in terms of the FIMP parameters.
We also express the warmness quantity in terms of the FIMP parameters.
We derive constraints on the FIMP parameters from several lower bounds on $m_{\mathrm{WDM}}$.
By using the analytic formulas, we also check the validity of deriving the constraints from the warmness quantity, by comparing the result with that derived from the linear matter power spectra.
Since the model shares common features with a broad class of FIMP models (2-body decay and $2 \to 2$ scattering), one can take advantage of this model to infer a (rough) constraint on other FIMP models.

This paper is organized as follows.
In the next section, we introduce our Benchmark FIMP model.
We provide the analytic formula of the phase space distribution.
In Sec.~\ref{sec:constr-from-warmess}, we introduce our warmness quantity and provide its analytic formula in terms of the FIMP model parameters.
We derive constraints on the FIMP model parameters from the reported lower bounds on $m_{\mathrm{WDM}}$.
We devote Sec.~\ref{sec:summary-discussion} to a summary and discussion.
In Appendix~\ref{sec:constr-from-line}, we derive the constraints on the FIMP model parameters by comparing the linear matter power spectra and compare them with those obtained from the warmness quantity.

\section{Setup}
\label{sec:setup}

\subsection{Benchmark FIMP model}
\label{sec:bench-fimp-model}

We consider Majorana fermion DM $\chi$, which feebly interacts with a Dirac fermion $\Psi$ and a complex scalar $\phi$.
$\phi$ also couples to a Dirac fermion $f$ in the thermal plasma.%
\footnote{
This FIMP model virtually corresponds to the light axino model considered in Refs.~\cite{Bae:2017tqn, Bae:2017dpt}.
The axino FIMP model is based on a supersymmetric version of Dine-Fischler-Srednicki-Zhitnitsky axion model~\cite{Zhitnitsky:1980tq, Dine:1981rt}.
Axino is a fermionic supersymmetric partner of axion that dynamically explains why the strong interaction preserves $CP$ very precisely~\cite{Peccei:1977hh, Peccei:1977ur, Weinberg:1977ma, Wilczek:1977pj}.
One can identify $\chi$, $\Psi$, $\phi$, and $f$, respectively, as light axino, Higgsino (supersymmetric partner of Higgs), Higgs, and top quark in the axino FIMP model.
}
The Lagrangian relevant to freeze-in production of $\chi$ is 
\begin{align}
  \label{eq:lagrangian}
  \mathcal{L}_{\mathrm{F.I.}}
  &=
-y_\chi \phi\bar\Psi \chi - y_f \phi\bar{f} f + \hc \,,
\end{align}
where $y_\chi$ and $y_f$ are Yukawa couplings.
We assume that $\Psi$ is in equilibrium with the thermal plasma.
In the following, we consider the mass spectrum of $m_\chi, m_f \ll m_\phi < m_\Psi$.%
\footnote{
The opposite spectrum, $m_\phi > m_\Psi$, makes scattering process irrelevant in the following discussion, but does not change other results drastically~\cite{Bae:2017dpt}.
}

Due to the feeble interaction with the other particles, $y_\chi \ll 1$, $\chi$ is not in equilibrium with the thermal plasma.
Meanwhile $\chi$ is produced by freeze-in processes:
\begin{itemize}
\item
  2-body decay: $\Psi \to \chi \phi$, $\bar\Psi \to \chi \phi^*$
\item
  $t$-channel scattering: $\Psi f \to \chi f$, $\Psi \bar{f} \to \chi \bar{f}$, $\bar\Psi f \to \chi f$, $\bar\Psi \bar{f} \to \chi \bar{f}$
\item
  $s$-channel scattering: $f\bar{f} \to\chi\Psi$,  $f\bar{f} \to\chi\bar\Psi$
\end{itemize}
The scattering processes are mediated by $\phi$.
The freeze-in production is most efficient when the heaviest particle in the process becomes non-relativistic.
After that the production is suppressed by the Boltzmann factor.
Thus we define the decoupling temperature $T_{\mathrm{dec}} = m_\Psi$.


\subsection{Phase space distribution}
\label{sec:phase-space-dist}

We define the DM phase space distribution such that the DM total number density is given by
\begin{align}
  \label{eq:nchi}
  n_\chi(t) = g_\chi \int \frac{d^3p}{(2\pi)^3} f_\chi (t, p)\,,
\end{align}
where $t$ is the cosmic time, $p$ is the physical momentum, and $g_\chi=2$ is the DM spin degrees of freedom.
The phase space distribution follows the Boltzmann equation.
It is generically very challenging to solve the Boltzmann equation.
On the other hand, due to the feeble interaction, one can linearize the Boltzmann equation in terms of $f_\chi$.

First, since $\chi$ free-streams after the production, its momentum is just redshifted.
Thus it is convenient to describe the phase space distribution as a function of $q = p/T_\chi$.
The temperature of $\chi$ is defined by
\begin{align}
  \label{eq:T-chi}
  T_\chi = \left(\frac{g_{*s}(T)}{g_{*s}(T_{\mathrm{dec}})}\right)^{\frac{1}{3}}T,
\end{align}
where $g_{*s}$ is the number of effective massless degrees of freedom for entropy and $T$ is the plasma temperature.

Second, in the collision term, one can set $f_\chi \simeq 0$.
Thus, the collision term is reduced to a sum of the collision term of each production process.
One can obtain the resultant phase space distribution from each production process as
\begin{align}
  \label{eq:phase-space-general}
  g_\chi f_\chi(q)
  &=
    \int_{t_i}^{t_f}dt\, \frac{g_\chi}{E_\chi}C(t,p)
    \simeq
    \frac{1}{H(T_{\mathrm{dec}})}
    \int_{x_i}^{x_f}dx\, x \frac{g_\chi}{E_\chi}C(t,p)\,,
\end{align}
where $x=m_\Psi/T$.
$C(t,p)$ is the collision term of that process with $f_\chi \simeq 0$ (see Ref.~\cite{Bae:2017dpt} for explicit expressions).
For other particles, we neglect the Pauli-blocking and Bose-enhancement effects, $1-f \simeq 1$, and approximate the thermal distribution by the Maxwell-Boltzmann distribution: $f^{\mathrm{eq}}\simeq e^{-E/T}$.%
 \footnote{The first approximation slightly changes the peak position of $q^2f(q)$, but the effect is at most $\Order(1)\,\%$~\cite{Bae:2017dpt}. The second approximation only affects the overall yield.}
$H(T_{\mathrm{dec}})$ is the expansion rate at $T=T_{\mathrm{dec}}=m_\Psi$:
\begin{align}
  \label{eq:hubble}
  H(T_{\mathrm{dec}}) = \sqrt{ \frac{\pi^2}{90} g_*(T_{\mathrm{dec}})}\frac{m_\Psi^2}{M_{\mathrm{pl}}} \equiv \frac{m_\Psi^2}{M_0}\,,
\end{align}
with the reduced Planck mass $M_{\mathrm{pl}}$. 
We ignore the temperature dependence in the number of effective massless degrees of freedom for energy $g_*(T)$ and instead use a constant $g_*(T_{\mathrm{dec}})$ since the freeze-in production is most efficient at $T\sim T_{\mathrm{dec}} = m_\Psi$.
For the same reason, we set $x_i = 0$ and $x_f = \infty$.

Now one can perform the time integral in Eq.~\eqref{eq:phase-space-general} analytically.
Here are the resultant phase space distributions from each production process:
\begin{align}
  g_\chi f_{\text{2-body}}(q)
  &=
    \frac{y_\chi^2M_0}{4\sqrt{\pi} m_\Psi}\times
    (1-r^2)^{-1}
    \left(\frac{q}{1-r^2}\right)^{-\frac{1}{2}}
    \exp\left(-\frac{q}{1-r^2}\right)\,,
    \label{eq:dist-2body-1-anal}
  \\
  g_\chi f_{t\text{-ch}}(q)  
  &=
     \frac{y_\chi^2y_f^2M_0}{16\pi^{\frac{5}{2}} m_\Psi}
    \times q^{-\frac{1}{2}}e^{-q} \times
    \frac{(2-r^2)\tanh^{-1}\sqrt{1-r^2} -\sqrt{1-r^2}}{3(1-r^2)^{\frac{3}{2}}}\,,
    \label{eq:dist-tch-1-anal}
  \\
  g_\chi f_{s\text{-ch}}(q)  
  &=
    \frac{y_\chi^2y_f^2M_0}{16\pi^{\frac{5}{2}} m_\Psi}
    \times q^{-\frac{1}{2}}e^{-q}
    \notag\\
  &\times
    \frac{
    \pi \left(2+(-3+2q)r^2+r^4\right)
    \mathrm{Erfc}\left(\sqrt{\frac{q}{1-r^2}}\right)
    \exp\left(\frac{q}{1-r^2}\right)
    -2\sqrt{\pi}r^2\sqrt{1-r^2}q^{\frac{1}{2}}
    }{4(1-r^2)^{5/2}}\,,
    \label{eq:dist-sch-1-anal}
\end{align}
where $r = m_\phi / m_\Psi$ and $\mathrm{Erfc}(x)$ is the complementary error function defined by
\begin{align}
  \label{eq:erfc-def}
  \mathrm{Erfc}(x)
  &=
    \frac{2}{\sqrt \pi}\int_{x}^\infty dt\,e^{-t^2}\,.
\end{align}

\begin{figure}
  \centering
  \begin{minipage}{0.33\linewidth}
    \includegraphics[width=1.0\linewidth]{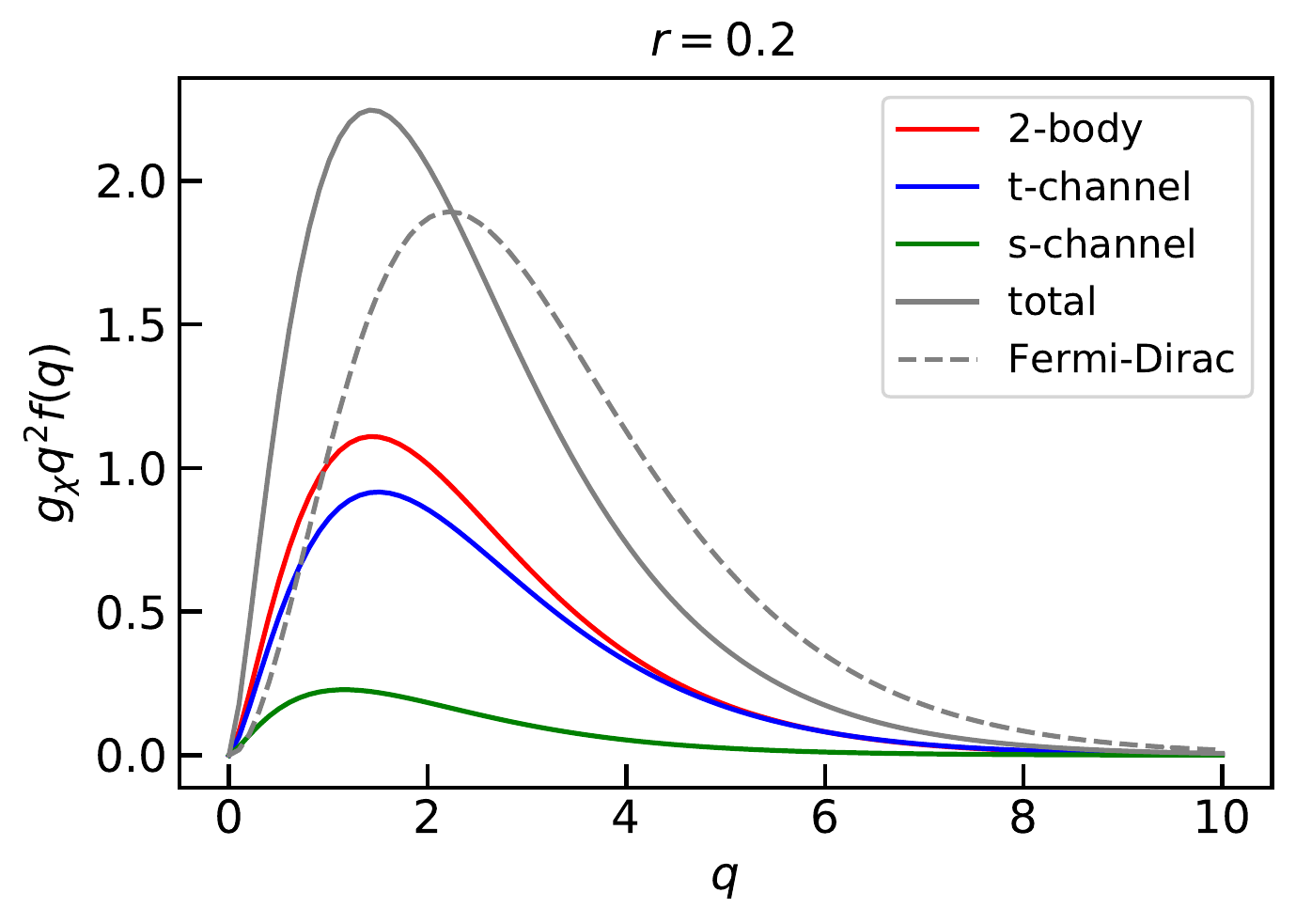}
  \end{minipage}%
   \begin{minipage}{0.33\linewidth}
    \includegraphics[width=1.0\linewidth]{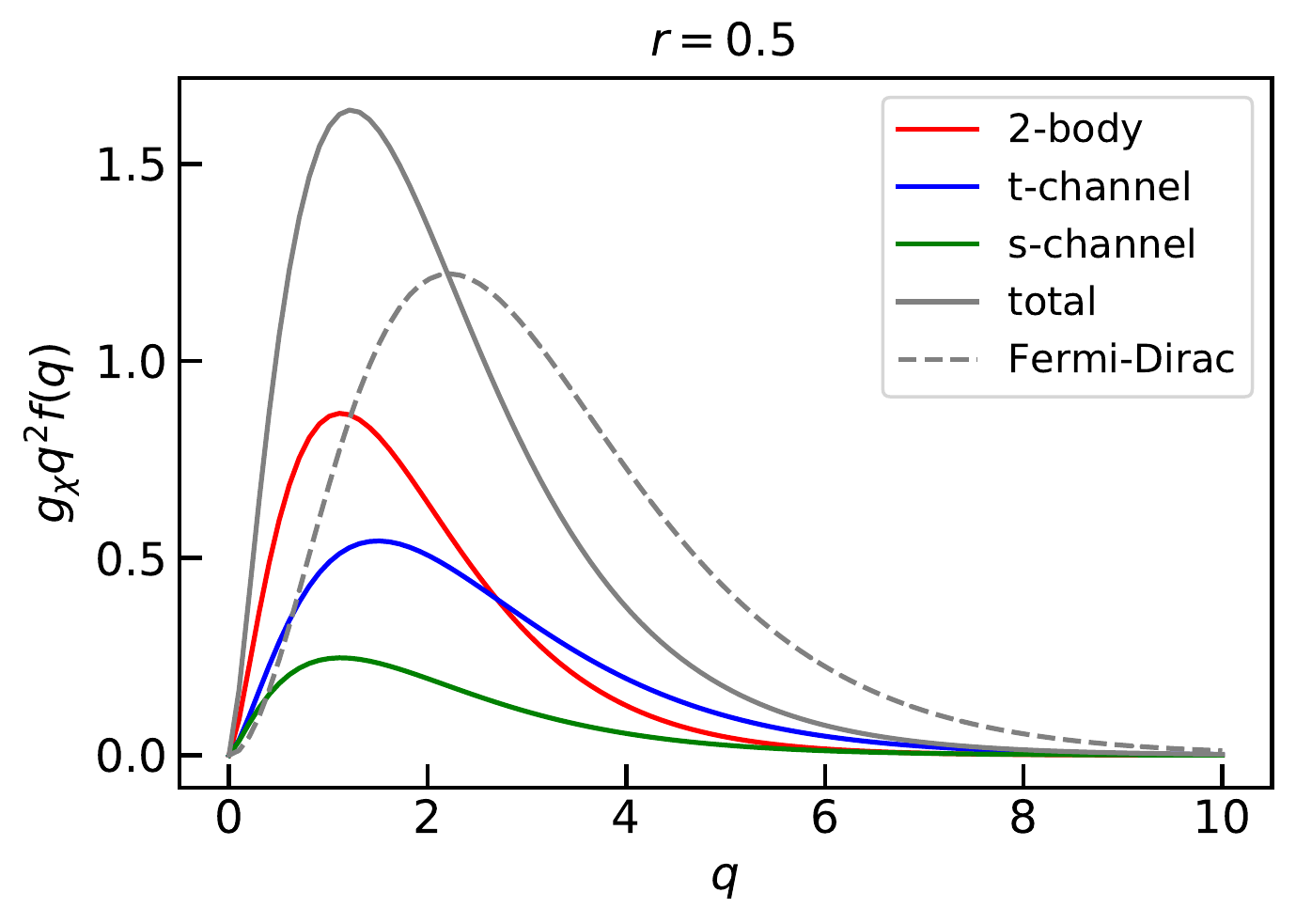}
  \end{minipage}%
   \begin{minipage}{0.33\linewidth}
    \includegraphics[width=1.0\linewidth]{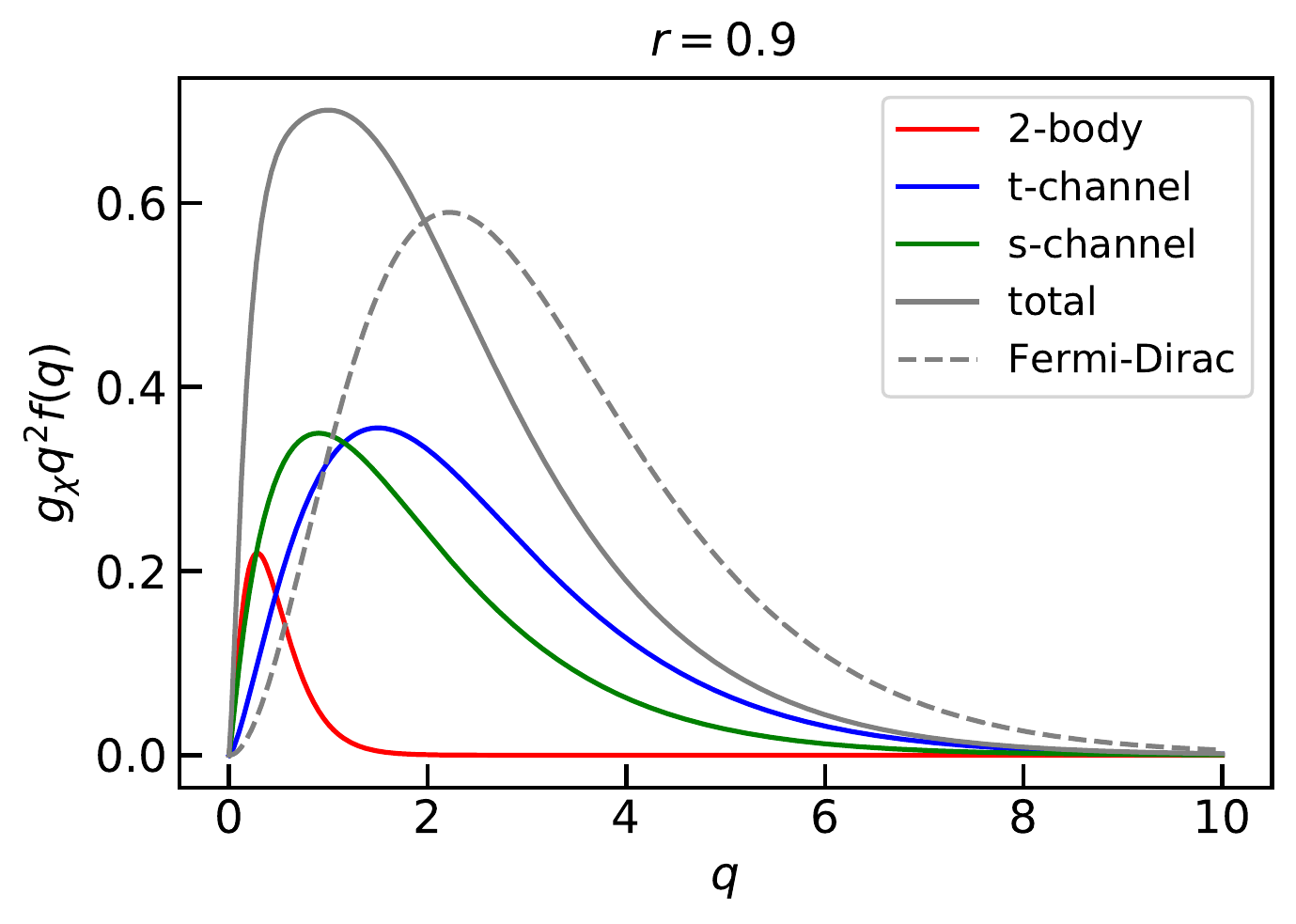}
  \end{minipage}
  \caption{Phase space distributions from 2-body decay (red), $t$-channel scattering (blue) and $s$-channel scattering (green) for $r=0.2$ (left), $0.5$ (middle) and $0.9$ (right). The total distribution is shown by gray solid line. Other parameters are fixed as $y_f=\sqrt{4\pi}$, $y_\chi=10^{-7.5}$ and $m_\Psi = 10^{-16}M_0$. The number of particle species ($\Psi/{\bar \Psi}$ and $f/\bar f$) are also taken into account. For comparison, the Fermi-Dirac distribution, $f(q) \propto 1 /(e^{q} + 1)$ normalized such that the yield ($\int dq q^2 f(q)$) is identical to the FIMP total yield, is shown by gray dashed line.}
  \label{fig:phase-space-dist}
\end{figure}
The above formula of the phase space distribution from the 2-body decay is consistent with that in Refs.~\cite{Heeck:2017xbu, Boulebnane:2017fxw}.
For the 2-body decay, the phase space distribution is sensitive to the mass degeneracy: for $r \to 1$, it becomes very cold (a large population at a low momentum), while the yield is suppressed by the factor of $(1-r^2)^2$.
On the other hand, for the $t$-channel scattering, the mass degeneracy affects only the yield slightly: the shape and thus warmness of the phase space distribution is independent of $r$.
For the $s$-channel scattering, $r$-dependence is complicated: the phase space distribution becomes slightly colder for $r \to 1$.
These scattering processes play a role when $r$ is close to unity and $y_f$ is $\Order(1)$.
We show sample spectra for $r=0.2, 0.5$ and $0.9$ in Fig.~\ref{fig:phase-space-dist}.

\section{Constraining light FIMP DM from a warmness quantity}
\label{sec:constr-from-warmess}

The constraints on WDM from the structure formation of the Universe are often reported in the conventional thermal WDM.
Like neutrinos, the thermal WDM particles follow the Fermi-Dirac distribution with the spin degrees of freedom being 2: $f_{\mathrm{WDM}}(t,p) = 1/(e^{p/T} + 1)$ and $g_{\mathrm{WDM}} = 2$, and hence the relic abundance is parametrized by its temperature $T_{\mathrm{WDM}}$ and mass $m_{\mathrm{WDM}}$ as
\begin{align}
  \label{eq:thermal-wdm-abundance}
  \Omega_{\mathrm{WDM}}h^2
  =
  \left( \frac{m_{\mathrm{WDM}}}{94\unit{eV}}\right)
  \left( \frac{T_{\mathrm{WDM}}}{T_\nu} \right)^3
  =
  7.5
  \left( \frac{m_{\mathrm{WDM}}}{7\unit{keV}}\right)
  \left( \frac{106.75}{g_{*s,\mathrm{WDM}}} \right)\,.
\end{align}
For a given thermal WDM mass $m_{\mathrm{WDM}}$, $T_{\mathrm{WDM}}$ is fixed such that the thermal relic coincides with the observed DM abundance.
In the second equality, we have used Eq.~\eqref{eq:T-chi} and $g_{*s,\mathrm{WDM}} = 106.75$ counts all the SM degrees of freedom.
Note that $g_{*s,\mathrm{WDM}}\sim 7000$ is required for $m_{\mathrm{WDM}} \sim 7 \keV$ to achieve $\Omega_{\mathrm{WDM}} h^2 \sim 0.1$ and thus some entropy production is implicitly assumed.
In this work, among the reported constraints from Lyman-$\alpha$ forest observations, we use $m_{\mathrm{WDM}} > 5.3 \keV$~\cite{Irsic:2017ixq} as a stringent bound and $m_{\mathrm{WDM}} > 2.0 \keV$~\cite{Viel:2005qj} as a conservative bound.

In order to constrain the light FIMP DM, in principle, we need to repeat the process of
\begin{itemize}
\item[]
\begin{center}
Model $\to$ DM phase space distribution $\to$ Linear matter power spectrum $\to$ Observables.
\end{center}
\end{itemize}
It is very easy to obtain the linear matter power spectrum from the phase space distribution (second step).
One can incorporate Eqs.~\eqref{eq:dist-2body-1-anal}-\eqref{eq:dist-sch-1-anal} into a Boltzmann solver like \texttt{CLASS}~\cite{Blas:2011rf, Lesgourgues:2011rh} in a straight forward manner.
But calculating observables still requires hard and time-consuming efforts.
Instead, in this paper, we map the reported constraints on $m_{\mathrm{WDM}}$ onto the FIMP parameters.

\subsection{Warmness quantity}
\label{sec:warmness}

We introduce the following warmness quantity of DM, which is calculable from the phase space distribution~\cite{Kamada:2013sh}:
\begin{align}
  \label{eq:warmness}
  \sigma = \frac{\sqrt{\vev{p^2}}}{m_\mathrm{DM}} = \tilde{\sigma}  \frac{T_\mathrm{DM}}{m_\mathrm{DM}}\,,
\end{align}
where
\begin{align}
  \label{eq:sigmatilde}
  \tilde\sigma^2 =
  \frac{\int d^3q\, q^2 f(q)}{\int d^3q\,f(q)}\,.
\end{align}
This characterizes the sound speed and thus the Jeans scale of DM~\cite{Kamada:2013sh}.

We can construct a map between FIMP parameters and thermal WDM mass $m_{\mathrm{WDM}}$ by equating the warmness: $\sigma_{\chi} = \sigma_{\mathrm{WDM}}$. 
The temperature of $\chi$ is given by Eq.~\eqref{eq:T-chi}, while the temperature of WDM (equivalently $g_{*s, \mathrm{WDM}}$) is fixed to reproduce the observed DM abundance by Eq.~\eqref{eq:thermal-wdm-abundance}. Then we obtain
\begin{align}
  \label{eq:m-chi-limit}
  m_\chi = 7\keV \left(\frac{m_{\mathrm{WDM}}}{2.5\keV}\right)^{\frac{4}{3}}\left(\frac{\tilde\sigma_\chi}{3.6}\right)\left(\frac{106.75}{g_{*s} (T_{\mathrm{dec}})}\right)^{\frac{1}{3}}\,,
\end{align}
where the shape of the FIMP phase space distribution, $f(q)$ (see Eqs.~\eqref{eq:dist-2body-1-anal}-\eqref{eq:dist-sch-1-anal} and Fig.~\ref{fig:phase-space-dist}), is imprinted through $\tilde\sigma_\chi$.
Note that the thermal WDM, where WDM particles follow the Fermi-Dirac distribution, has $\tilde{\sigma}_{\mathrm{WDM}} \simeq 3.6$.
The net $\tilde\sigma_\chi$ is a yield-weighted sum of the warmness of the phase space distribution from each production process:
\begin{align}
  \label{eq:sigmatile-total}
  \tilde\sigma^2_\chi
  =
  \frac{Y_{\chi,\text{2-body}}}{Y_{\chi}}\tilde\sigma^2_{\chi,\text{2-body}}
  +
  \frac{Y_{\chi,t\text{-ch}}}{Y_{\chi}}\tilde\sigma^2_{\chi,t\text{-ch}}
  +
  \frac{Y_{\chi,s\text{-ch}}}{Y_{\chi}}\tilde\sigma^2_{\chi,s\text{-ch}}\,,
\end{align}
where in the present model, each $\tilde\sigma^2$ is calculated from Eqs.~\eqref{eq:dist-2body-1-anal}-\eqref{eq:dist-sch-1-anal} analytically
\begin{align}
  \tilde\sigma^2_{\chi,\text{2-body}}
  &=
  \frac{35}{4}(1-r^2)^2\,,
  \label{eq:sigmatilde-2body-1}
  \\
  \tilde\sigma^2_{\chi,t\text{-ch}}
  &=
  \frac{35}{4}\,,
  \label{eq:sigmatilde-tch}
  \\
  \tilde\sigma^2_{\chi,s\text{-ch}}
  &=
    \frac{7(105r-265r^3+191r^5-15r^7-15(1-r^2)^3(7+r^2)\tanh^{-1}r)}{12r^4(r(3-r^2)+(-3+2r^2+r^4)\tanh^{-1}r)}\,.
  \label{eq:sigmatilde-sch}
\end{align}
The DM yield (number density per entropy) is also analytically calculated as
\begin{align}
  Y_{\chi,\text{2-body}}
  &\simeq 2\times\frac{3y_\chi^2 M_1}{32\pi^2 g_{*s}(T_{\mathrm{dec}}) m_\Psi}
  \left( 1 -r^2 \right)^2\,,
  \label{eq:yield-2body-anal}\\
  Y_{\chi,t\text{-ch}}
  &\simeq
  4\times \frac{3 y_\chi^2y_f^2 M_1}{128\pi^4 g_{*s}(T_{\mathrm{dec}}) m_\Psi}\times
  \frac{(2-r^2)\tanh^{-1}\sqrt{1-r^2} -\sqrt{1-r^2}}{3(1-r^2)^{\frac{3}{2}}}\,,
  \label{eq:yield-tch-anal}\\
  Y_{\chi,s\text{-ch}}
  &\simeq
  2\times \frac{3 y_\chi^2y_f^2M_1}{128\pi^4 g_{*s}(T_{\mathrm{dec}}) m_\Psi}\times
  \frac{r(3-r^2)+(-3+2r^2+r^4)\tanh^{-1}(r)}{2r^5}\,,
  \label{eq:yield-s-anal}
\end{align}
where $M_1 = 45 / (2 \pi^2) M_0$ with $M_0$ given by Eq.~\eqref{eq:hubble}.
Prefactors count the number of particle species ($\Psi/{\bar \Psi}$ and $f/\bar f$).

One may wonder to what extent this map works well.
In Appendix~A of Ref.~\cite{Bae:2019sby}, the constraints obtained from direct modeling are compared with those derived from the warmness quantity.
They agree with each other up to $\sim 10\%$ in the lower bound on the FIMP mass.
Furthermore, we compare the constraints obtained here with those derived from the linear matter power spectra in Appendix~\ref{sec:constr-from-line}.
Again we find the maximum difference is up to $\sim 10\%$.

\subsection{Obtained constraints}
\label{sec:constr}

\begin{figure}
  \centering
  \begin{minipage}{0.49\linewidth}
    \includegraphics[width=1.0\linewidth]{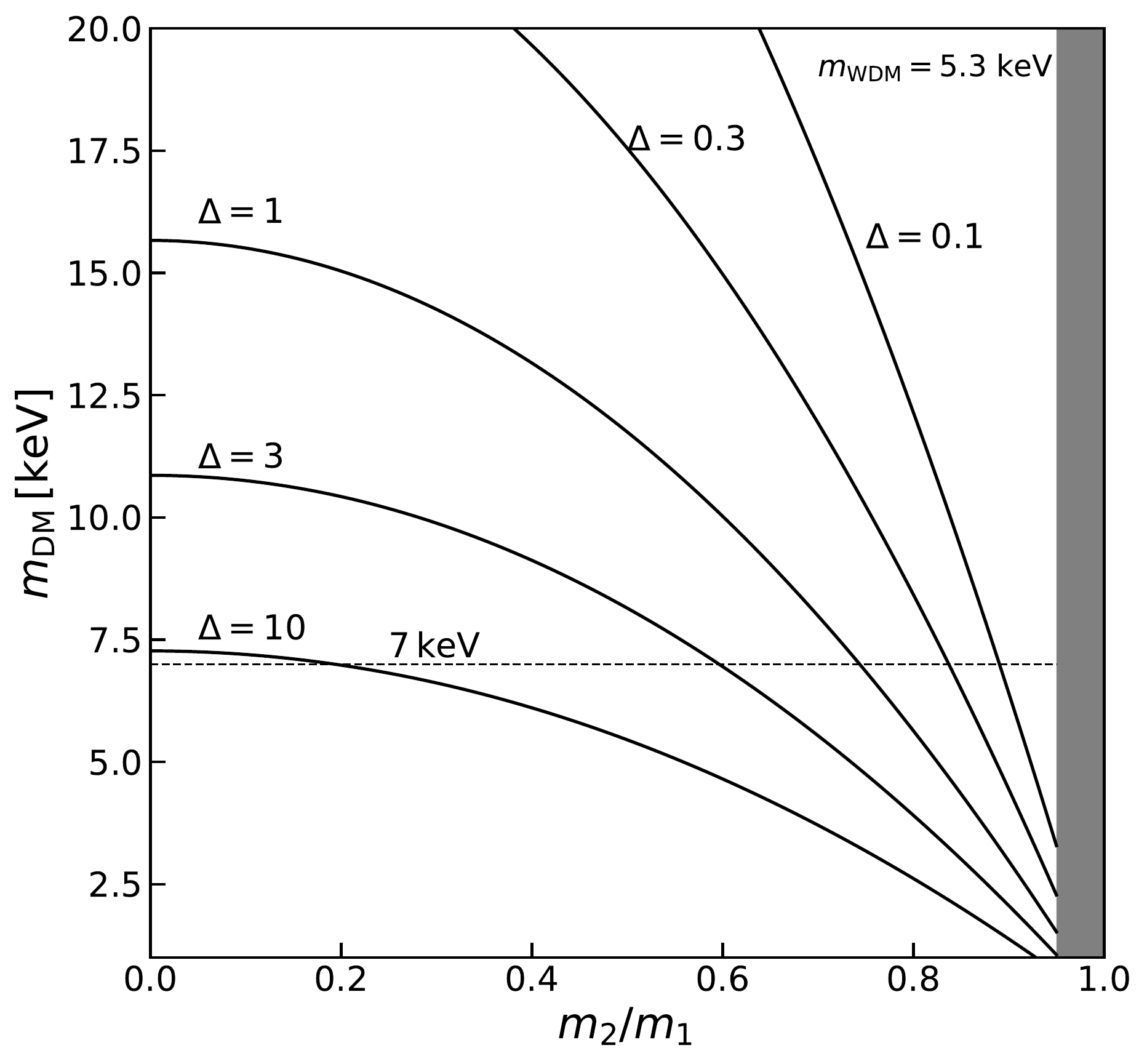}
  \end{minipage}%
  \begin{minipage}{0.49\linewidth}
    \includegraphics[width=1.0\linewidth]{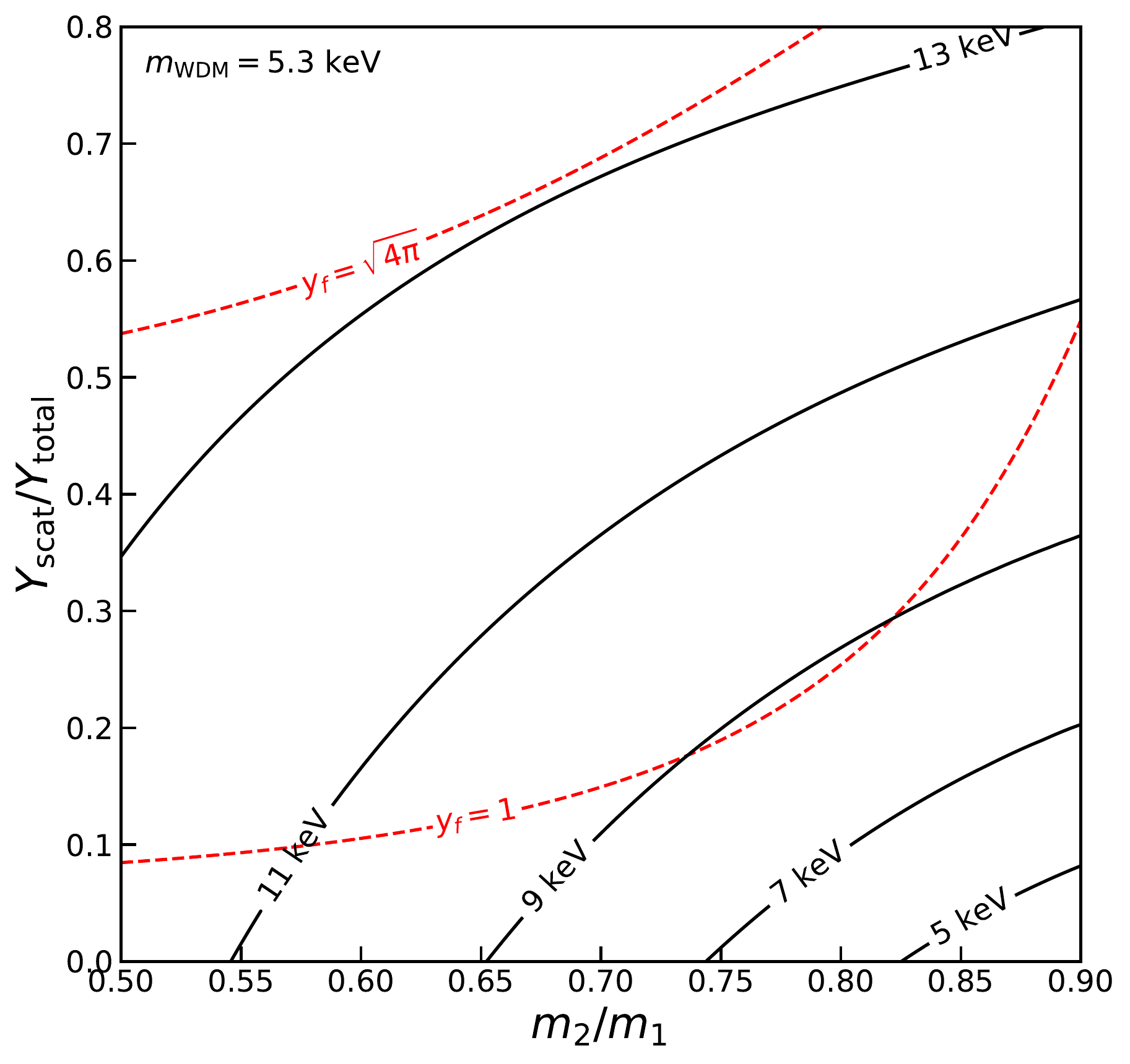}
  \end{minipage}
  \begin{minipage}{0.49\linewidth}
    \includegraphics[width=1.0\linewidth]{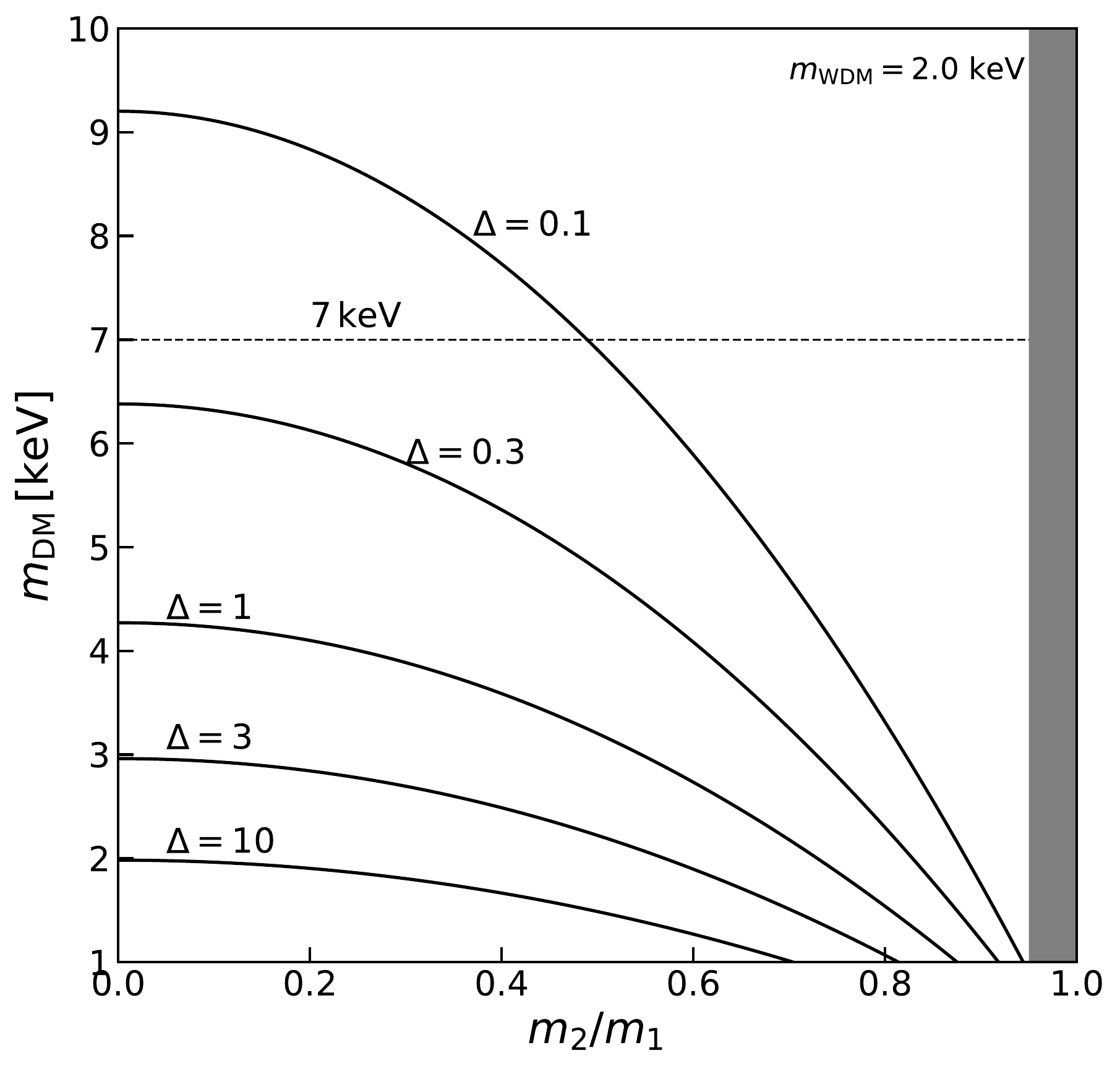}
  \end{minipage}%
  \begin{minipage}{0.49\linewidth}
    \includegraphics[width=1.0\linewidth]{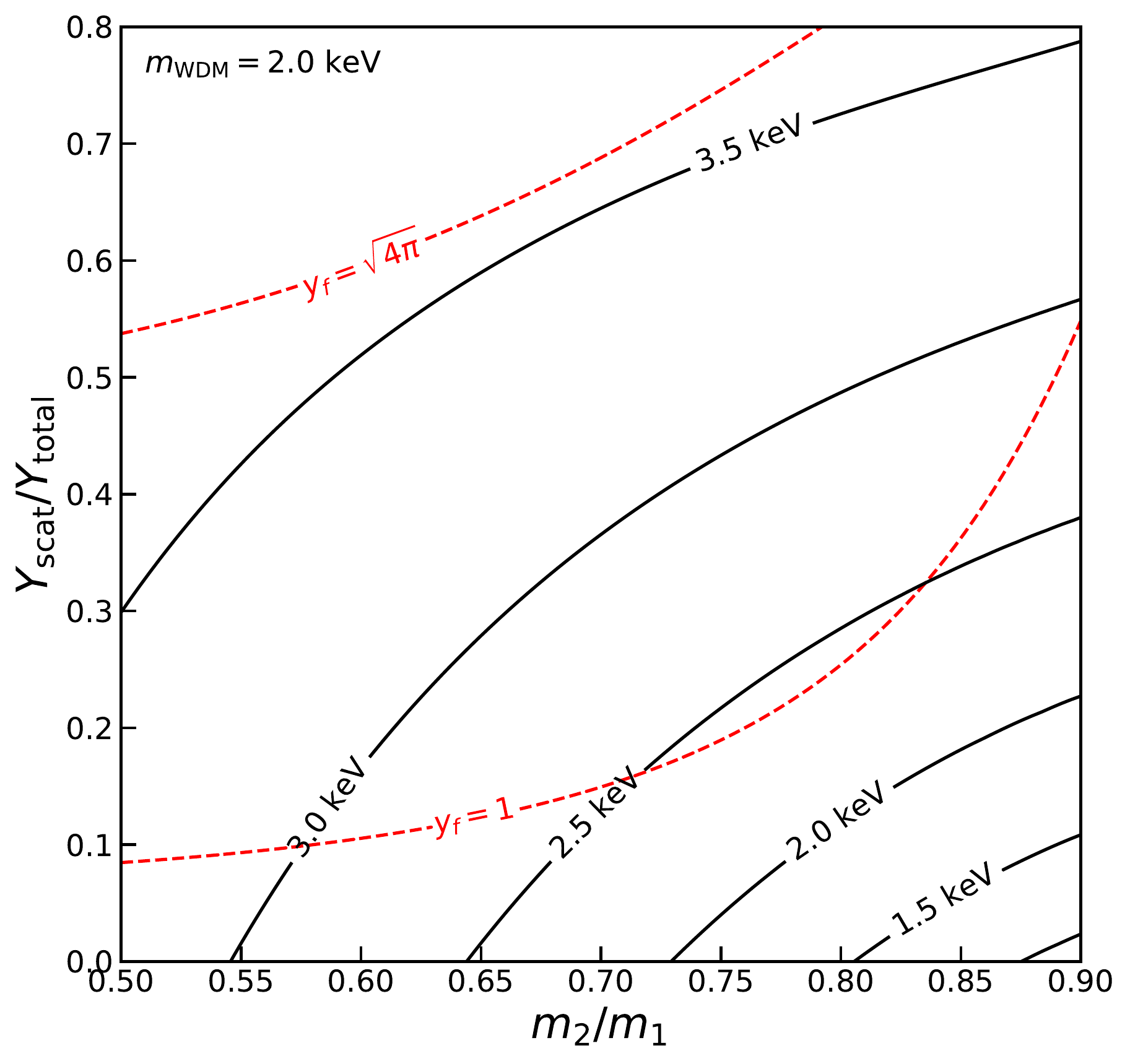}
  \end{minipage}
  \caption{Constraints from the map constructed from $\tilde{\sigma}$, which is given by Eq.~\eqref{eq:m-chi-limit}. We use the notation of $m_{\mathrm{DM}} = m_{\chi}$, $m_1 = m_\Psi$, and $m_2 = m_\phi$. Left panels are the constraints on the FIMP from 2-body decay. For each $\Delta$, the region below the line is disfavored. Right panels are constraints on 2-body decay + scattering. At each model point $(m_2/m_1, Y_{\mathrm{scat}}/Y_{\mathrm{total}})$, the lower bound on FIMP mass is obtained, indicated by the black solid contour. Upper panels show the constraints corresponding to $m_{\mathrm{WDM}}=5.3\keV$, and lower panels show those corresponding to $m_{\mathrm{WDM}}=2\keV$.
Red dashed counters correspond to $y_\chi=1$ and $\sqrt{4\pi}$.}
  \label{fig:sigma}
\end{figure}

Now it is straightforward to derive a constraint on the FIMP parameters with the help of the map given by Eq.~\eqref{eq:m-chi-limit}.
In the top (bottom) two panels of Fig.~\ref{fig:sigma}, we show the Lyman-$\alpha$ forest constraints corresponding to $m_{\mathrm{WDM}} > 5.3\keV$ ($m_{\mathrm{WDM}}>2.0\keV$).
In the figure, we use the notation of $m_{\mathrm{DM}} = m_{\chi}$, $m_1 = m_\Psi$, and $m_2 = m_\phi$ for a general use.
The gray-shaded regions of the left panels, corresponding to $r > 0.95$, is not included since thermal effects may be important for such a degenerate spectrum~\cite{Hamaguchi:2011jy}.

In the left two panels, we assume that FIMP DM is produced predominantly by the 2-body decay.
We use $g_{*s} (T_{\mathrm{dec}}) = 106.75 \times \Delta$, where $\Delta$ parametrizes the amount of the entropy production.
For example, $\Delta = 1$ (0.1) is taken if FIMP DM is produced most efficiently around the electroweak phase transition (neutrino decoupling) and no entropy production occurs later.
For each $\Delta$, the region below the line is disfavored by the Lyman-$\alpha$ forest constraints.
In particular, we see that the latest lower bound, $m_{\mathrm{WDM}} > 5.3\keV$, disfavors $7\keV$ FIMP DM produced by freeze-in of the 2-body decay, unless entropy production occurs after the production or a degenerate mass spectrum ($m_2/m_1 \gtrsim 0.75$) is taken.

In such a degenerate mass spectrum, the decay width of $\Psi \to \phi \chi$ is so suppressed that the scattering contribution to the yield may become relevant.
As we mentioned around Eqs.~\eqref{eq:dist-2body-1-anal}-\eqref{eq:dist-sch-1-anal}, the phase space distribution from the scatterings does not get that cold as $m_2/m_1 = r \to 1$ in contrast to that from the 2-body decay.
In the right two panels, we show the constraints on FIMP DM produced by 2-body decay + scattering, while fixing $\Delta = 1$.
We use the notation of $Y_{\mathrm{scat}} = Y_{\chi,t\text{-ch}} + Y_{\chi,s\text{-ch}}$ and $Y_{\mathrm{total}} = Y_{\chi,\text{2-body}} + Y_{\chi,t\text{-ch}} + Y_{\chi,s\text{-ch}}$.
For each model parameter $(m_2/m_1, Y_{\mathrm{scat}}/Y_{\mathrm{total}})$, the lower bound on FIMP mass is obtained, indicated by the solid contour.
As long as the scattering contribution is negligible, ($Y_{\mathrm{scat}}/Y_{\mathrm{total}} \simeq 0$), $7 \keV$ FIMP DM with $m_2/m_1 > 0.75$ is allowed even by the Lyman-$\alpha$ forest constraint corresponding to $m_{\mathrm{WDM}}>5.3\keV$ (top panel).
Once the scattering contribution exceeds about $10\,\%$ of the total DM yield, FIMP DM becomes hotter and is disfavored by the Lyman-$\alpha$ forest constraint corresponding to $m_{\mathrm{WDM}}>5.3\keV$ (top panel).
This demonstrates the general trade-off between the FIMP warmness and yield: FIMP DM gets colder for a more degenerate mass spectrum of particles involved in the decay channel; but at some point, the scattering contribution to the yield becomes relevant and FIMP DM gets hotter for a further more degenerate mass spectrum.
FIMP DM cannot be arbitrarily cold.

\section{Summary and discussion}
\label{sec:summary-discussion}

A light (keV-scale) FIMP is an intriguing alternative to a WIMP as a DM candidate:
it is sufficiently long-lived without any symmetry and its rare decay is detectable in x-ray observations;
and it behaves as WDM in contrast to CDM and alters the galactic-scale structure formation.
It is essential to combine on-going and in-coming x-ray observations and measurements of the galactic-scale structure.
On the other hand, it is not straightforward to derive a constraint on FIMP parameters from the structure formation.
One has to repeat the following hard procedure on a model-by-model basis:
\begin{itemize}
\item[]
\begin{center}
Model $\to$ DM phase space distribution $\to$ Linear matter power spectrum $\to$ Observables.
\end{center}
\end{itemize}

To improve the situation, in this paper, we have introduced a benchmark model of FIMP DM.
A big advantage of this FIMP model is that the analytic formula of the phase space distribution of FIMPs is available (first step).
It is easy to incorporate our analytic formula into a public Boltzmann solver like \texttt{CLASS} and thus obtain the linear matter power spectrum (second step).
Therefore in this benchmark model, we can save the first two steps to derive constraints from observations.
Our benchmark model will facilitate FIMP searches in the structure formation of the Universe.

Moreover, we can take a further simplification to derive a constraint: introducing a certain warmness quantity.
We have developed a map between the thermal WDM mass $m_{\mathrm{WDM}}$ and the FIMP parameters by equating the warmness quantity.
By using this map, one can derive constrains on the FIMP parameters from the reported lower bounds on $m_{\mathrm{WDM}}$.
We have indeed derived a constraint from the latest Lyman-$\alpha$ forest data ($m_{\mathrm{WDM}} > 5.3 \keV$).
Our results indicate that $7 \keV$ FIMP DM, without entropy production or a degenerate spectrum, is in tension with the latest Lyman-$\alpha$ forest data.

Our analytic map will be very useful when another x-ray line signal is found and/or constraints from the structure formation are updated in future.
One can just adopt our analytic map in our benchmark FIMP model as an approximation of his/her own FIMP model.
Then, it is very easy to check (or more precisely infer) whether or not a FIMP DM explanation to the signal is compatible with the structure formation of the Universe.
This does not cause a big difference from the above direct modeling, unless one does not need a precise value (likely the case in particle physics model-building).
We also note that in such a case, one opts for direct modeling, but needs to control (typically larger) systematic errors of modeling and astrophysical processes.
Our analytic map will facilitate FIMP model-building.

\section*{Acknowledgments}

We thank Kyu Jung Bae and Ryusuke Jinno for useful discussions.
The work of AK is supported by IBS under the project code, IBS-R018-D1.
AK would like to acknowledge the Mainz Institute for Theoretical Physics (MITP) of the Cluster of Excellence PRISMA+ (Project ID 39083149) for enabling AK to complete a significant portion of this work.
The work of KY was supported by JSPS KAKENHI Grant Number JP18J10202.

\appendix

\section{Warmness quantity v.s. linear matter power spectrum}
\label{sec:constr-from-line}

The constraints in Sec.~\ref{sec:constr} are obtained from a map given by Eq.~\eqref{eq:m-chi-limit} between $m_{\mathrm{WDM}}$ and the FIMP parameters.
This map is constructed from the warmness quantity given by Eq.~\eqref{eq:warmness}, which is calculable from the phase space distribution.
Here, we adopt another way to map the reported constraints on $m_{\mathrm{WDM}}$ onto the FIMP parameters.

It is based on the comparison of the linear matter power spectra.
For that purpose, we define the transfer function as
\begin{align}
  \label{eq:T2k}
  T^2(k) = \frac{P(k)}{P_{\mathrm{CDM}}(k)}\,,
\end{align}
and the half mode $k_{1/2}$ as
\begin{align}
\label{eq:khalf}
    T^2(k_{1/2}) = 1/2\,.
\end{align}
We regard a given FIMP model parameter set as disfavored if $k_{1/2, \chi} < k_{1/2,\mathrm{WDM}}$.

We incorporate the analytic formula of the phase space distribution \eqref{eq:dist-2body-1-anal}-\eqref{eq:dist-sch-1-anal} in \texttt{CLASS}, and compute the linear matter power spectrum at present with the cosmological parameters from “Planck 2015 TT, TE, EE+lowP” in Ref.~\cite{Ade:2015xua}.
Then we identify the half mode $k_{1/2}$ from the resultant linear matter power spectra.

\begin{figure}
  \centering
  \begin{minipage}{0.49\linewidth}
    \includegraphics[width=1.0\linewidth]{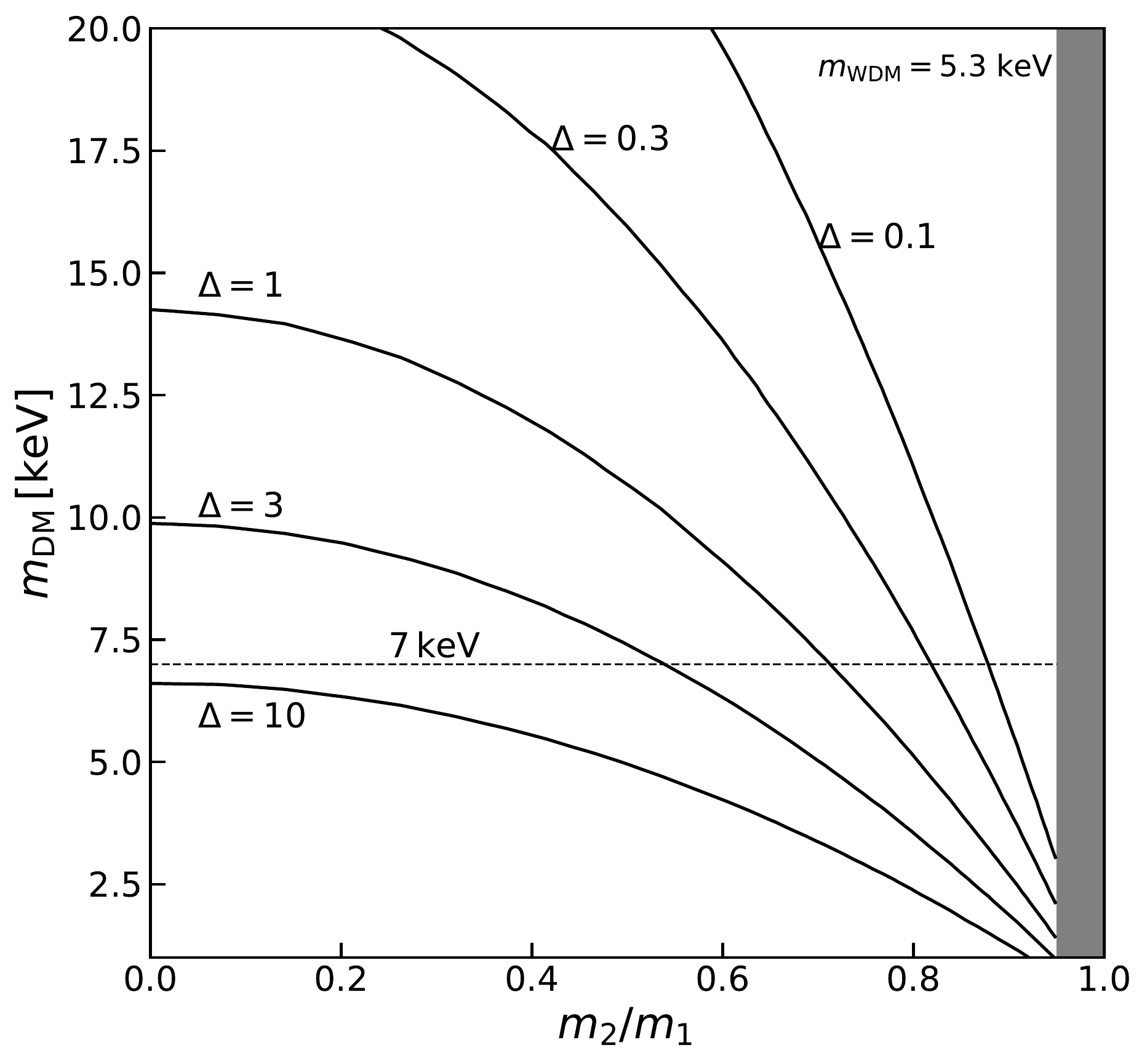}
  \end{minipage}%
  \begin{minipage}{0.49\linewidth}
    \includegraphics[width=1.0\linewidth]{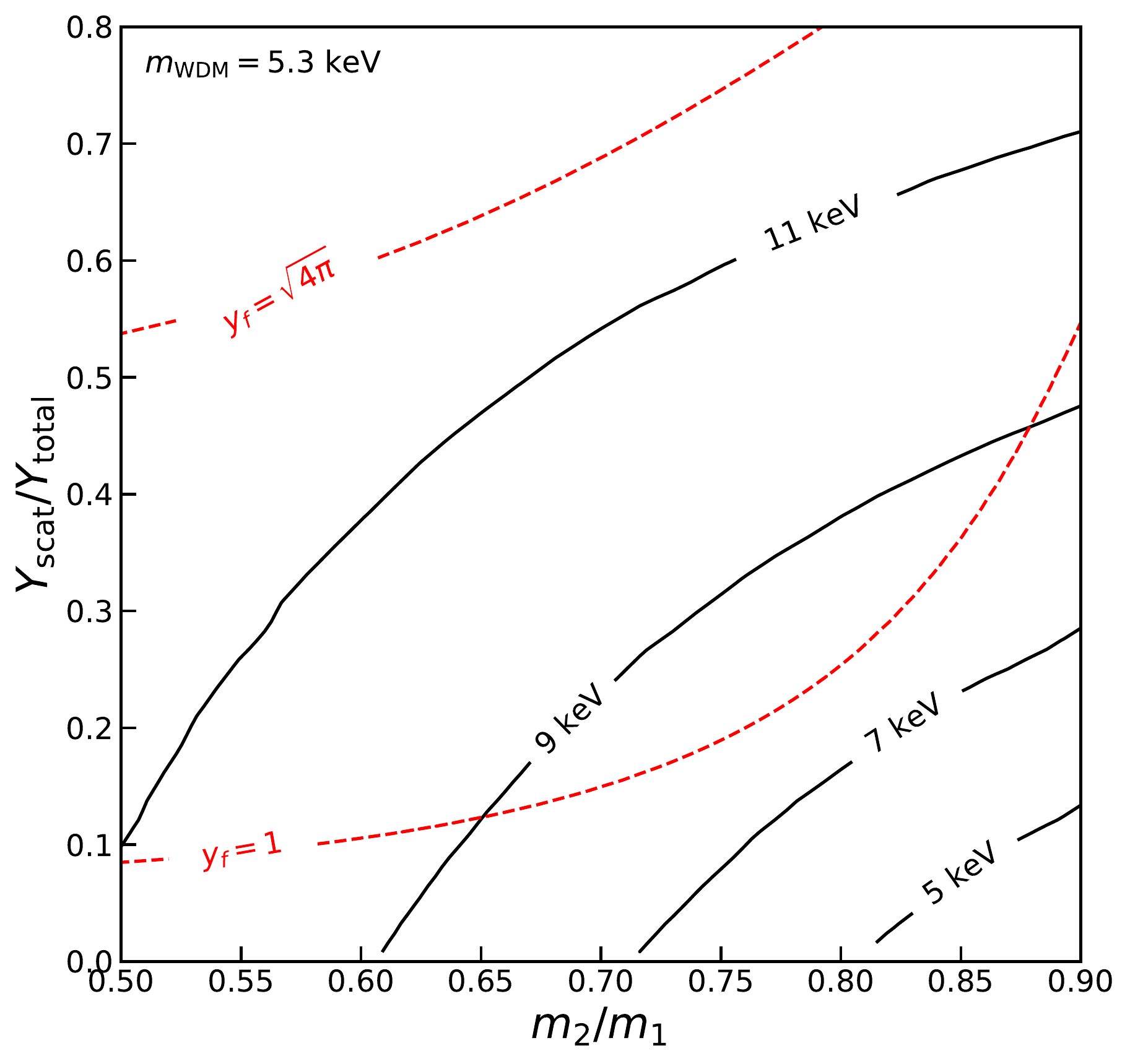}
  \end{minipage}
  \begin{minipage}{0.49\linewidth}
    \includegraphics[width=1.0\linewidth]{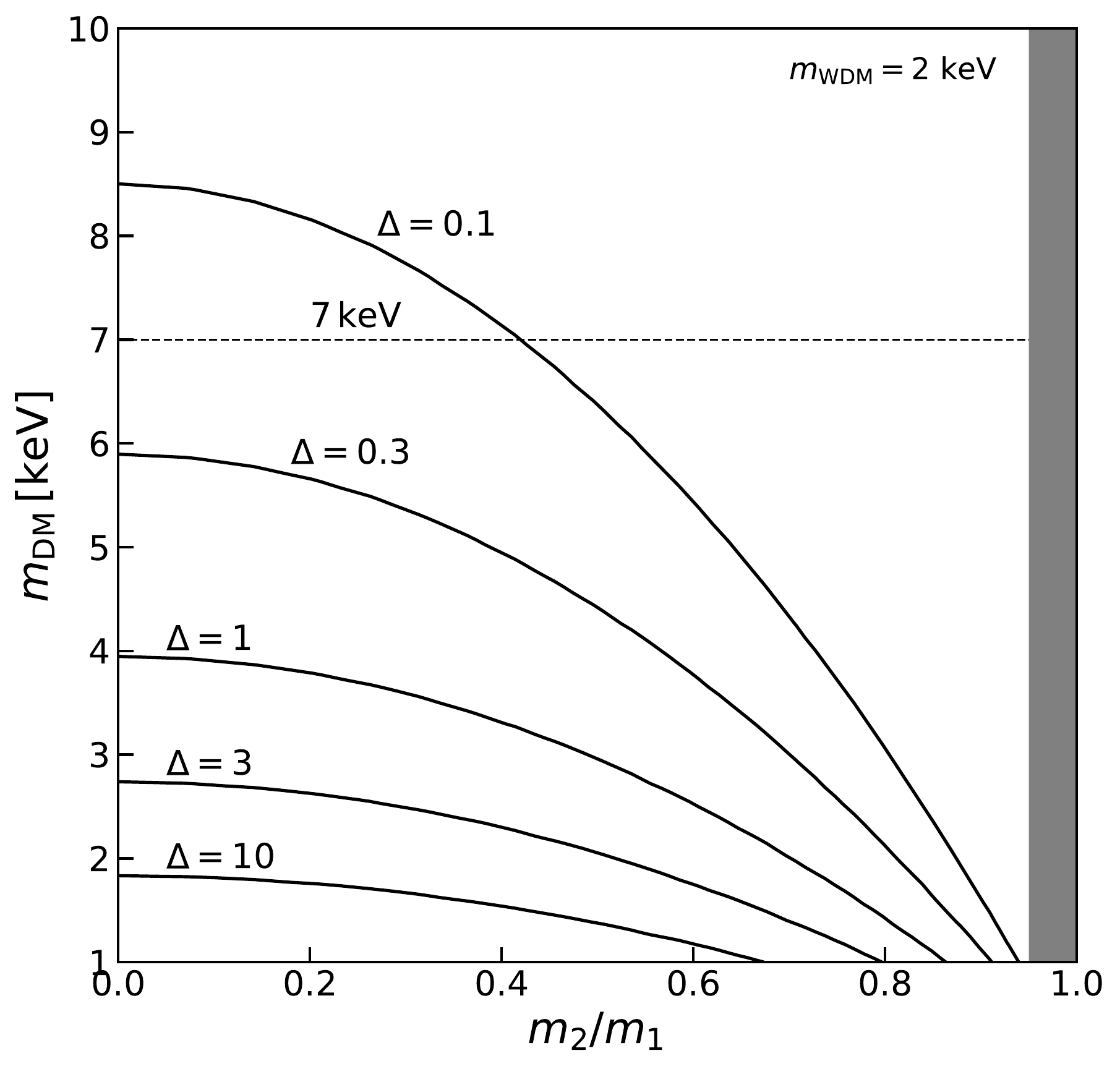}
  \end{minipage}%
  \begin{minipage}{0.49\linewidth}
    \includegraphics[width=1.0\linewidth]{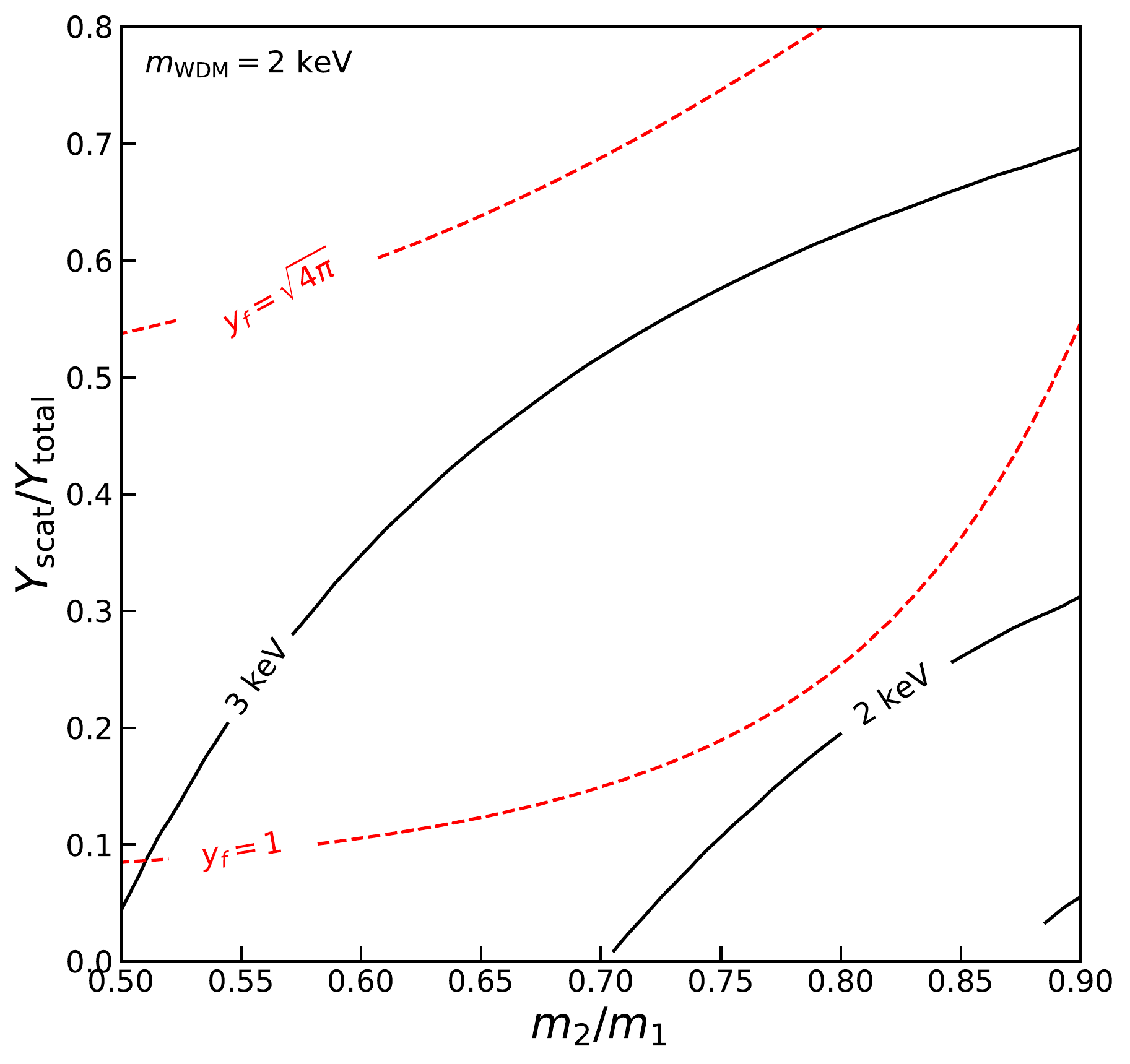}
  \end{minipage}
  \caption{Constraints from $k_{1/2}$. Each panel of this figure should be compared with the corresponding panel of Fig.~\ref{fig:sigma}.}
  \label{fig:khalf}
\end{figure}

Fig.~\ref{fig:khalf} shows the constraints obtained from the comparison of $k_{1/2}$. 
Each panel should be compared to the corresponding panel of Fig.~\ref{fig:sigma}.
We see that the derived bounds are in agreement with each other up to at maximum $10\%$ difference in $m_{\mathrm{DM}}$.

\bibliographystyle{utphysmod}
\bibliography{manuscript} 

\end{document}